\documentclass[prl,twocolumn,showpacs,superscriptaddress,amsmath,amssymb]{revtex4} 

\usepackage{dcolumn}   
\usepackage{bm}        
\usepackage{amssymb}   

\usepackage[T1]{fontenc}
\usepackage[latin1]{inputenc}
\usepackage[letterpaper]{geometry}
\usepackage[dvips]{graphicx} 
\usepackage{amsthm}
\usepackage{amstext}
\usepackage{amsmath}
\usepackage{graphicx}
\usepackage{amssymb}

\makeatletter

\providecommand{\be}{\begin{equation}}
\providecommand{\ee}{\end{equation}}

\providecommand{\beni}{\begin{equation*}}
\providecommand{\eeni}{\end{equation*}}

\theoremstyle{plain}

\makeatother

\begin{document}

\title{The Hierarchical Random Energy Model}
\pacs{05.10.-a,05.50.+q,75.10.Nr}

\author{Michele Castellana} 
\affiliation{Laboratoire de Physique Th\'eorique et Mod\`eles
  Statistiques, CNRS - Universit\'e Paris Sud, B\^at. 100, 91405 Orsay Cedex, France} 
\affiliation{Dipartimento di Fisica, Universit\`a di Roma `La Sapienza' , 00185 Rome, Italy}
\author{Aur\'elien Decelle} 
\affiliation{Laboratoire de Physique Th\'eorique et Mod\`eles Statistiques, CNRS - Universit\'e Paris Sud, B\^at. 100, 91405 Orsay Cedex, France} 
\author{Silvio Franz}
\affiliation{Laboratoire de Physique Th\'eorique et Mod\`eles Statistiques, CNRS - Universit\'e Paris Sud, B\^at. 100, 91405 Orsay Cedex, France} 
\author{Marc M\'ezard} 
\affiliation{Laboratoire de Physique Th\'eorique et Mod\`eles
  Statistiques, CNRS -  Universit\'e Paris Sud, B\^at. 100, 91405 Orsay Cedex, France} 
\author{Giorgio Parisi}
\affiliation{Dipartimento di Fisica, Universit\`a di Roma `La Sapienza' , 00185 Rome, Italy}

\begin{abstract} We introduce a Random Energy Model on a hierarchical
  lattice where the interaction strength between variables is a
  decreasing function of their mutual hierarchical distance, 
making it  a non-mean field model.  Through
  small coupling series expansion and a direct numerical solution of the
  model, we provide evidence for a spin glass condensation transition
  similar to the one occurring in  the usual mean field Random Energy Model. At variance with
  mean field, the high temperature branch of the free-energy is
  non-analytic at the transition point.
\end{abstract}

\maketitle

Clarifying the nature of glassy states is a fundamental goal of modern
statistical physics. Both for spin glasses \cite{MPV} and for structural glasses \cite{castellani-cavagna}, the
mean field theory of disordered systems provides a suggestive picture
of laboratory glassy phenomena as the reflection of an ideal thermodynamic
phase transition. Unfortunately, the development of a first principles theory of
glassy systems going beyond mean field, has resisted decades of
intense research \cite{6,7,8}.  One of the main obstacles towards this goal, lies
in the lack of reliable real space renormalization group (RG) schemes 
allowing to reduce the effective number of degrees of freedom and
identify the relevant fixed points describing glassy phases. In
ferromagnetic systems an important role in the understanding of the
real space RG transformation has been played by spin systems with
power law interactions on hierarchical lattices \cite{dyson, collet-eckmann}.  In these models the RG equations take the simple form of non-linear integral 
equations for an unknown function (as opposed to the 
functional of statistical
field theory), that can be solved with high precision. In this
perspective it  is natural to generalize these models
to  spin glasses \cite{SK-H}. \cite{foot}

In this letter, we introduce the simplest such spin glass
model, a random energy model (REM) \cite{derrida,gross-mezard}. As we shall see, the hierarchical REM 
is such that the interaction energy between subsystems scales subextensively
in the system size. It thus
qualifies as a non-mean-field model. We report in what follows the
results of a small
coupling expansion and of an algorithmic solution of the RG equations for
the entropy that, exploring complementary regions of parameter space,  provide the first analytic evidence in favor of an ideal glass transition in a non mean-field model.
Interestingly, this transition turns out to have -as in the case of the standard REM- the character of an entropy catastrophe analogous to the one hypothesized long ago for the 
structural glasses \cite{Kauzmann,Adam}. 

The hierarchical REM can be defined as a system of $N=2^k$
Ising spins with an energy function defined recursively.
The recursion is started at the level of a
single spin $k=0$, with the definition of $H_0[S]=\epsilon_0(S)$, where
the single spin energies are independent identically distributed (i.i.d.) random variables extracted from a distribution $\mu_0(\epsilon)$.  At the level $k+1$, we
consider then two independent systems of $2^{k}$ spins $S_1=\{S_{1i}\},\,
i=1,\ldots,2^{k}$ and $S_2=\{S_{2i}\},\, i=1,\ldots,2^{k}$ with Hamiltonians
$H_{1k}\left[S_{1}\right]$ and $H_{2k}\left[S_{2}\right]$ respectively
and put them in interaction to form a composite system of $2^{k+1}$
spins and Hamiltonian
\begin{eqnarray}\label{eq:107}
H_{k+1}\left[S_{1},S_{2}\right]&=&H_{1k}\left[S_{1}\right]+H_{2k}\left[S_{2}\right]+\epsilon_{k}\left[S_{1},S_{2}\right], \nonumber
\end{eqnarray}
where the $\epsilon_k$ are i.i.d. random variables extracted from a distribution $\mu_{k+1}(\epsilon)$, chosen to have 
zero mean and variance 
$\left<\epsilon_{k}\left[S_{1},S_{2}\right]^{2}\right>\backsim 2^{(k+1)(1-\sigma)}$.
The interaction term $\epsilon_{k}\left[S_{1},S_{2}\right]$ is
physically analogous to a surface interaction energy between the two
subsystems. For $\sigma\in (0,1)$ this model qualifies as a non mean-field system, 
where the interaction energy between
different parts of the system scales with volume to a power smaller
than unity. On the contrary, when $\sigma \leq 0$ the interaction
energy  grows faster than the volume.  A rescaling of the energy  is
then necessary to get 
a well defined thermodynamic limit. The system behaves in this case as a mean-field model.
Finally, for $\sigma> 1$ the interaction energy decreases with distance and asymptotically  the model behaves as a free system. In the following we focus on the most interesting region $0<\sigma<1$.

We have studied this model with two different methods. The first one
is a replica study of the quenched free-energy, performed through 
a small coupling perturbative expansion. The second one is a numerical
estimate of the microcanonical entropy as a function of the energy.
Both methods suggest that a REM-like finite-temperature phase transition occurs for all $\sigma\in (0,1)$. 

\textit{Perturbative computation of the free-energy}- In order to make the calculations as simple as possible, we have chosen a Gaussian distribution for the energies 
$\epsilon_k$.\\ 
We then considered the perturbative expansion in $g\equiv 2^{1-\sigma}$ of the  free-energy
$f(T)=f^{(m)}(T)+O(g^{m+1}).$
Notice that the expansion of $f$ to the $m$-th order takes into account just the interactions with range less or equal to $2^m$, i. e. the first $m$ hierarchical levels. 

The computation of the free-energy has been done with the replica
method. In this context it is just a mathematical tool to organize the terms of the series. We considered then the expansion of the average partition function of the system replicated $n$ times
\begin{equation}\label{eq:200}
\overline{Z^{n}}=\sum_{ S_{1}\cdots S_{n}}\exp\left[\frac{\beta^{2}}{4}\sum_{j=0}^{k}g^{j}\sum_{i=1}^{2^{k-j}}\sum_{a,b=1}^{n}\delta_{ S_{a}^{(j,i)} S_{b}^{(j,i)}}\right]
\end{equation}
where $\beta \equiv 1/T$,  and $S^{(j,i)}$ 
is the configuration of the 
$i$-th group of spins at the $j$-th level of the hierarchy. 
This representation allowed an automated computation of $f^{(m)}$ up to the value of $m=10$. 

A useful check of the method is obtained considering $\sigma<0$. Since
in this case high values of $j$ dominate the energy in (\ref{eq:200}), correlations
between the energy levels can be neglected. After rescaling the
energies by $\epsilon_j \rightarrow 2^{k\sigma /2}\epsilon_j$ the free-energy of the model becomes equal to the one of the standard REM \cite{derrida, gross-mezard} with 
critical temperature $T_c=\sqrt{\sum_j 2^{-\sigma j}/\log 2}$.
We found that, when increasing $m$, $f^{(m)}$ converges to the REM free-energy with exponential speed in the whole high temperature regime  $\beta<\beta_c$.

\begin{figure}[htb] 
\centering
\includegraphics[width=7.5cm]{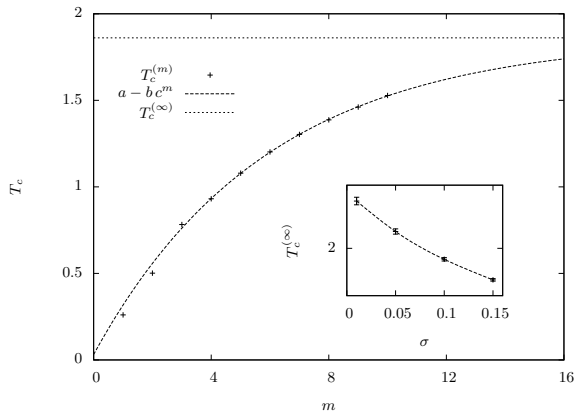}
\caption{The temperatures $T_c^{(m)}$ vs $m$ for $\sigma=.1$. Here $T_c=1.861\pm.021$. 
Inset: $T_c$ vs $\sigma$.}
\label{Flo:1}
\end{figure}

\begin{figure}[htb] 
\centering
\includegraphics[width=7.5cm]{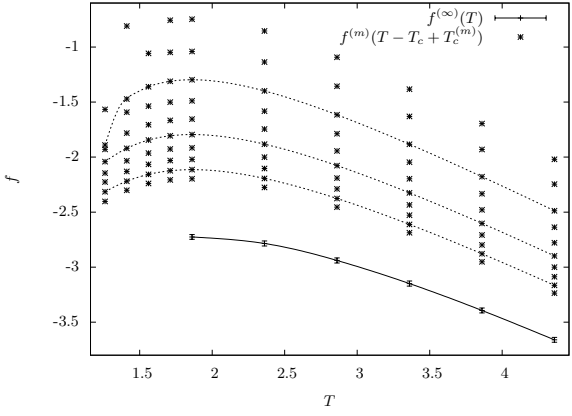}
\caption{
To get a better convergence for the free-energy, we considered the sequence $f^{(m)}(T-T_c+T_c^{(m)})$ instead of $f^{(m)}(T)$. As $T_c^{(m)}\rightarrow T_c$ for $m\rightarrow \infty$, the two sequences have the same limit $f^{(\infty)}(T)$. Here we see that for $\sigma=.1$, $f^{(m)}(T)$ has negative entropy $s^{(m)}(T)$ for $T<T_c^{(m)}$.
}
\label{Flo:9}
\end{figure}

We now consider $f(T)$ for $0<\sigma<1$. The direct inspection of the curves shows that,
for $m \ge 1$, the $ m$-th order entropy
$s^{(m)}(T) \equiv -{df^{(m)}(T)}/{dT}$ while positive at high
temperature, becomes negative at some temperature $T_{c}^{(m)}$.  As
can be seen in fig. \ref{Flo:1}, the sequence $T_{c}^{(m)}$
exhibits a good exponential convergence to a finite limit
$T_{c}$ for $\sigma \leq 0.15$. The stability of these data for large $m$ clearly
suggests that an entropy crisis transition is present in the model at $T_c$.
The inset in fig. (\ref{Flo:1}) shows that $T_c$ is a decreasing
function of $\sigma$, consistently with the fact that the larger
$\sigma$, the weaker the interaction strength. At high temperature also the free-energy series has a
good exponential convergence in $g$ (see fig. \ref{Flo:9}). 

This small $g$ expansion gives some evidence for an entropy crisis
taking place at temperature $T_c$.  It is important to realize that
this $T_c$ cannot be simply computed from the sum of the
variances of the $\epsilon_k$: the energy correlations cannot be neglected.
An entropy crisis implies the existence of  a phase transition at a temperature $\ge T_c$. In a REM
scenario, the phase transition would take place exactly at $T_c$, when
the entropy vanishes. An argument in favour of such a result can be
found with a one-step replica symmetry breaking Ansatz. Consider the
partition function (\ref{eq:200}) and suppose that the $n$ replicas
are grouped into $n/x$ groups, so that, for any two replicas $a,b$ in
the same group, $S_{a}^{(j,i)} =S_{b}^{(j,i)}$ for all $i,j$. Then perform again the small $g$
expansion, within this Ansatz. To each order $m$, 
this procedure gives a free-energy $f^{(m)}_x(T)=f^{(m)}(T/x)$. The
maximization over $x$ then gives $x=1 $ for $T>T_c^{(m)}$, and
$x=T/T_c^{(m)}$ for $T<T_c^{(m)}$. This result is in complete analogy
with the one found in the REM, so the above replica symmetry breaking
Ansatz predicts a REM-like transition at $T=T_c$. In order to get a
distinct evidence for this scenario, we have done some numerical study.

\textit{Numerical computation of the entropy}- We exploit the hierarchical structure of the model to compute the
microcanonical entropy $S_k(E)$. 
In order to make the computations as simple as possible, we have chosen for $\mu_k(\epsilon)$ the Binomial distribution \cite{drem,kabashima}; 
$\mu_k(\epsilon)=\frac{1}{2^{M_k}} {\small \left(\begin{array}{c}M_k\\\epsilon+\frac{M_k}{2}\end{array}\right)}$.
At the level $k$,  $M_k$ is the integer part of
$\gamma2^{k(1-\sigma)}$, to have the same scaling of the variance as
in the Gaussian model. The constant $\gamma$ is chosen so that 
for all the values of $\sigma $ studied, $\lfloor(\gamma2^{k(1-\sigma)})\rfloor / (\gamma2^{k(1-\sigma)}) \approx 1$ for every $k$.
Consider the disorder-dependent density of states for a sample $a$ : 
${\cal N}_k^a(E)=\sum_{S}\delta_{H_k(S),E}$. The recursion relation that defines the 
model's Hamiltonian implies that when two samples $a$ and $b$ at the level $k$ 
are merged to define a sample at the level $k+1$, the resulting density of states 
${\cal N}_{k+1}^c(E)$ satisfies: 
\begin{eqnarray}
{\cal N}_{k+1}^c(E)=\sum_{E_a,E_b,\epsilon \atop 
E=E_a+E_b+\epsilon} n_k(E_a,E_b,\epsilon)
\label{NN}
\end{eqnarray}
where $n_k(E_a,E_b,\epsilon)=\sum_{S_1,S_2}\delta_{H_k^a(S_1),E_a}\times
\times \delta_{H_k^b(S_1),E_b}$ $\delta_{\epsilon_k(S_1,S_e),\epsilon}$ is the number of states in the composite
system that have $H_a=E_a$, $H_b=E_b$ and interaction energy equal to
$\epsilon$.  For given $E_a$ and $E_b$, the joint distribution of the
$n_k(E_a,E_b,\epsilon)$ for the different values of $\epsilon$ is
multinomial with parameters 
$q_\epsilon=\langle n_k(E_a,E_b,\epsilon)\rangle=
{\cal N}_k^a(E_a){\cal N}_k^b(E_b)\mu_k(\epsilon)$, while $n_k$'s with
different first or second argument are independent. 

\begin{figure}[htb] 
\centering
\includegraphics[width=7.5cm]{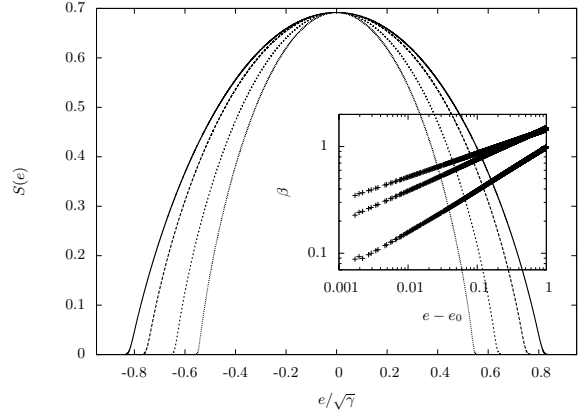}
\caption{The entropy $s(e)$ vs $e/\sqrt{\gamma}$ for $\sigma=.9,.8,.7,.6$ (with $\gamma=30,10,5,5$ respectively), from the outside to the inside. Inset: power law behaviour of $\beta-\beta_c$. The slopes are close to $1-\sigma$, with $\sigma =.6,.7,.8$ from bottom to top.}
\label{EntrFig}
\end{figure}

Our algorithmic approach starts from the exact iteration of
Eq. \ref{NN}. Thanks to the use of a discrete interaction energy, the
iteration time grows with $k$ proportionally to
$2^{k(3-\sigma)}$. This allowed us to reach the level $k=12$. The
results of the iteration shows that the values of the energy can be
divided in \textit{bulk region} of energy density around the origin
where the number of states ${\cal N}_k(E)= e^{2^k S_k(E/2^k)}$ is
exponential in the system size, and an \textit{edge region} where
the number of states is of order one (see fig.  \ref{EntrFig}).

In order to proceed further, we assume the existence and
self-averaging property of the entropy density $S(e)$ in the
thermodynamic limit. We then coarse-grain our description.  We
discretize the {\it energy density} in the bulk region and use an
approximated iteration for the entropy, where the sum (\ref{NN}) is
approximated by its maximum term. We account for the edge region using
the exact recursion for the $N_0=10,000$ lowest energy levels.  We can
in this way iterate many times and obtain a good estimate of the
thermodynamic limit behaviour.

In fig. \ref{EntrFig}, we present the
average  entropy density as a function of
the energy density $e$ for various values of $\sigma$.
In order to identify the
transition,  it is more convenient to average the data
obtained with a fixed energy difference from the fluctuating ground states. We can get in this way
good estimates of the value of the inverse critical temperature of the model
$\beta_c = s'(e_0)$.
An interesting feature emerging from our analysis is that close to the
ground state energy density $e_0$ the entropy is not analytic and behaves as $ \label{220} S(e) \approx \beta_c (e-e_0) + C (e-e_0)^{a}$ with $a$ well fitted by the value $a=2-\sigma$. 
This behavior, when translated in the canonical formalism, implies a singularity of the
free-energy close to $T_c$, $F(T)=E_0+Const\times(T-T_c)^{\frac{2-\sigma}{1-\sigma}}$, corresponding to a specific heat exponent $\alpha=-\frac{\sigma}{1-\sigma}$. 

\begin{figure}[htb] 
\centering
\includegraphics[width=7.5cm]{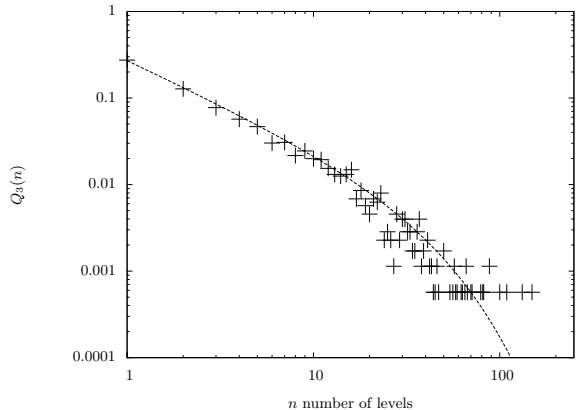}
\caption{Numerical data (cross) and the fitting function $Q_3(n)$  for the statistics of occupation of the ground state and the first two occupied levels. Here $k=10$, $\sigma=0.6$ and $\gamma=5$. The dashed line is a fit with the form (\ref{eq:220}) with $\beta_c = 1.20$}
\label{PoiFig}
\end{figure}

\begin{figure}[htb] 
\centering
\includegraphics[width=7.5cm]{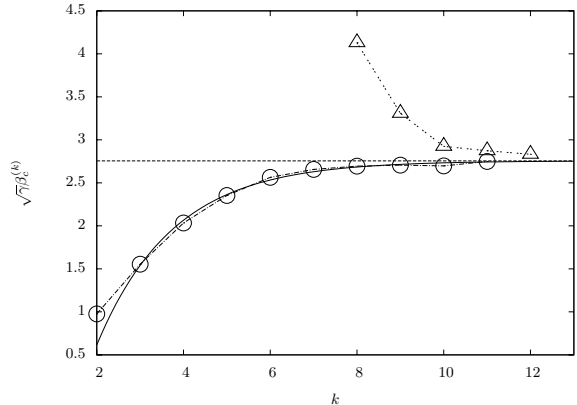}
\caption{Finite volume estimation of the inverse critical temperature, $\sqrt{\gamma} \beta_c^{(k)}$ vs $k$, for $\sigma=0.6$ and $\gamma=5$ determined by (\ref{eq:220}) (circle) and by 
$\beta_c = s'(e_0)$ (triangles). 
The latter have been fitted with a function of the type $\beta_c^{(k)}=\beta_c - B\, C^k$ (solid line). The asymptotic value $\beta_c$ is given by the dashed line.}
\label{ResFig}
\end{figure}

Having found evidence for a thermodynamic phase transition, we turn
our attention to the distribution of
low-lying energy states. The REM
picture suggests that, close to the ground state, the number of
energy levels with given energy $E$ are independent Poissonian
variables with density $\langle {\cal N}_\infty(E)\rangle =
e^{\beta_c(E-E_0)}$. A computation using extreme value statistics shows that
the probability $Q_\ell(k)$ that the ground state and first $\ell-1$ excited states are occupied by $n$ levels is given by: 
\begin{equation}\label{eq:220}
Q_\ell(n)	 = (1-\exp(-\ell \beta_c))^n /(\ell \beta_c n).
\end{equation}
In fig. \ref{PoiFig} we show the $Q_\ell(n)$ obtained numerically  together with 
a fit with the form (\ref{eq:220}). This procedure confirms the
validity of a REM-like transition, and provides  an alternative way of estimating the critical temperature.  As fig. \ref{ResFig} shows, the two estimates for different values of $k$ tend to the same limit from opposite directions.  

\textit{Conclusions}- In this letter we have introduced a hierarchical, non-mean field, 
REM. We have analyzed it through small coupling
series, 
through a 1RSB replica Ansatz,  and through an algorithmic approach. 
The two approaches point to the existence of a REM-like phase transition
at the temperature
where the entropy vanishes. At variance with the mean field result
(which predicts a discontinuity in the specific heat), one finds a
non-trivial specific heat exponent at $T_c$.
It will be interesting to study the replica structure of this
hierarchical REM in order to explore other possible replica solutions
at low temperatures.
 Another  important theme of future research is 
the study of spin-glass models with
 $p$-body interaction \cite{derrida,gardner,9}: at the mean
field level, these models display an entropy crisis  transition similar to the one of the REM
 whenever $p\geq 3$. It will be interesting to study them on hierarchical lattices.


\begin{thebibliography}{10}
\bibitem{foot}
These models should not be confused with models with local interactions on hierarchial lattices built on diamond plaquettes  \cite{Berker},  which, starting from ref. \cite{McKay} have been widely studied in their spin glass version, and lead to weakly frustrated systems even in their mean-field limit \cite{gardner2}.
\end{thebibliography}
\end{document}